\documentclass[apj]{emulateapj}
\usepackage{color,amsmath,amssymb,latexsym,txfonts}
\usepackage{graphicx,subfigure,threeparttable}
\usepackage{epstopdf}
\begin{document}
\title{Anomalous distributions of primary cosmic rays as evidence for time-dependent particle acceleration in Supernova remnants}
\author{Yiran Zhang$^{1,2}$, Siming Liu$^{1,2}$, Qiang Yuan$^{1,2}$}
\affil{$^1$Key Laboratory of Dark Matter and Space Astronomy, Purple
Mountain Observatory, Chinese Academy of Sciences, Nanjing 210008, China;
liusm@pmo.ac.cn (SL)\\
$^2$School of Astronomy and Space Science, University of Science and 
Technology of China, Hefei 230026, Anhui, China}
\begin{abstract}
Recent precise measurements of cosmic ray (CR) spectra show that the 
energy distribution of protons is softer than those of heavier nuclei, 
and there are spectral hardenings for all nuclear compositions above 
$\sim$200 GV. Models proposed for these anomalies generally assume 
steady-state solutions of the particle acceleration process. 
We show that, if the diffusion coefficient has a weak dependence on the 
particle rigidity near shock fronts of supernova remnants (SNRs), time-dependent solutions of the linear diffusive shock acceleration at two stages of SNR evolution can naturally 
account for these anomalies. The high-energy component of CRs is dominated by acceleration in the free expansion and adiabatic phases with enriched heavy elements and a high shock speed. The low energy component may be attributed to acceleration by slow shocks propagating in dense molecular clouds with low metallicity in the radiative phase. 
Instead of a single power-law distribution, the spectra of time-dependent 
solutions soften gradually with the increase of energy, which may be
responsible for the ``knee'' of CRs.
\end{abstract}
\section{Introduction}\label{sec1}
Supernova remnants (SNRs) have been considered as dominant sources of cosmic rays (CRs), especially for those with energies below the spectral ``knee'' at $\sim 10^{15}$ eV, the so-called Galactic CRs for their presumed Milky Way origin \citep{Hillas05,Ohira16}. There is also compelling observational evidence for efficient particle acceleration in SNRs \citep{Helder2012}. However, due to deflection of charged CRs by magnetic fields in the interstellar medium, propagation of CRs from their source regions to the Earth is not well-understood and it is still challenging to connect observational characteristics of SNRs to properties of CRs directly. It is generally accepted that CR sources inject a harder (broken) power-law distribution of high-energy particles into the Galaxy, which then softens to the spectrum observed near the Earth due to an energy-dependent diffusion process \citep{2012ApJ...761..133Y}. 

The mechanism of diffusive shock acceleration has been proposed for producing power-law high-energy particle distributions in SNRs \citep{1983RPPh...46..973D}. The test particle model predicts that in the steady-state case, high-energy particles in the shock downstream follow a power-law distribution with the index determined by the shock compression ratio. Microscopic details of the particle diffusion process only affect the upstream particle distribution and the time needed to reach the steady-state. However, for strong shocks of SNRs, the spectral index is close to 2, leading to a strong rigidity dependence of the escape rate of CRs from the Galaxy, which thus gives large anisotropies of the arrival directions of high energy CRs, in conflict with observations \citep{Hillas05,2017PrPNP..94..184A}. Multi-wavelength observations also do not support a single power-law particle distribution in SNRs \citep{Helder2012,2017ApJ...834..153Z,2017JHEAp..13...17O}.

Significant progresses have been made during the past decade. In particular, high precision CR flux measurements from a few GeV to a few TeV with the Alpha Magnetic Spectrometer (AMS) reveal several anomalies: 1) the spectrum of protons is softer than those of helium, carbon, and oxygen, with the spectral index different by $\gamma _{\textrm{\scriptsize p/He}}=-0.077\pm 0.002(\textrm{fit})\pm 0.007(\textrm{sys}) $ above the particle rigidity of 45 GV \citep{2015PhRvL.115u1101A}; 2) there is a spectral hardening at a transition rigidity of $ 336_{-44}^{+68}(\textrm{fit})_{-28}^{+66}(\textrm{sys}) $ GV for protons \citep{2015PhRvL.114q1103A} and $ 245_{-31}^{+35}(\textrm{fit})_{-30}^{+33}(\textrm{sys}) $ GV for helium \citep{2015PhRvL.115u1101A}. Similar results were also obtained by previous balloon and satellite experiments \citep{2009BRASP..73..564P,2010ApJ...714L..89A,2011Sci...332...69A}. These anomalies have been the subject of extensive studies and many models have been proposed \citep[see][for a review]{Ohira16}. In general, these anomalies can be attributed either to some propagation effects or to properties of CR sources. For the latter, the instantaneous distribution of accelerated particles has been assumed to be a power-law, which is appropriate if the particle acceleration timescale in the relevant energy range is much shorter than dynamical time of the accelerators.

However, the acceleration timescale of the highest energy particles in SNRs should be comparable to their ages in early stages of SNR evolution \citep{Helder2012}, and the gradual hardening of the radio spectral index with age, which challenges the steady-state approach of conventional diffusive shock models \citep{Reynolds2012}, also suggests that radio emitting electrons may be accelerated continuously during the evolution of SNRs \citep{2017ApJ...834..153Z}. Therefore the steady-state assumption may not be valid. Although AMS observations of secondary (boron) to primary (carbon) flux ratio reveals a power-law distribution with an index of $ \Delta =-0.333\pm 0.014(\rm fit)\pm 0.005(\rm sys) $ above 65 GV, implying a power-law rigidity dependence of the diffusion coefficient in the Galaxy with an index of 1/3 at the corresponding rigidities \citep{2016PhRvL.117w1102A}, such a scaling may not be valid near strong shocks of SNRs, where the diffusion can be dominated by turbulent mixture \citep{1993PhyU...36.1020B} and the acceleration rate can be suppressed significantly \citep{2010MNRAS.406.1337F, 2016arXiv161202262Y, 2016arXiv160908671H}. Moreover, if the threshold velocity for diffusive shock acceleration to operate is proportional to the shock speed, or considering differences in  acceleration of different ion species at low energies \citep{2004ApJ...610..550P}, the threshold rigidity of protons can be lower than other heavy elements \citep{2010ApJ...708..965Z}. One then expects a proton spectrum softer than other ions in the time-dependent solution as observed by the AMS.

A single component time-dependent solution of linear diffusive shock acceleration usually gives a gradually softening distribution at higher energies, which can not account for the spectral hardening above $\sim 200$ GV \citep[however, see][for an alternative]{Khialietal17}. The spectral hardening can be attributed to effects of different source populations, CR propagation, or non-linear acceleration of particles \citep{2012ApJ...752...68V, Ohira16}. In particular, the tentative detection of a spectral hardening in the lithium spectrum by AMS may hint at a propagation effect \citep{2012PhRvL.109f1101B}. On the other hand, a ``two-component'' model was also proposed by \citet{2015ApJ...815L...1T} for the spectral anomalies without considering details of the particle acceleration process. Here we adopt a smilar strategy of the ``two component'' model, but within the framework of time-dependent particle acceleration. Our model and results are given in \S\ \ref{sec2} and \S\ \ref{sec3}, respectively. In \S\ \ref{sec4} we draw conclusion and discuss the model implications.
\section{Model}\label{sec2}
For the sake of simplicity, we study the diffusive shock acceleration in SNRs by solving the one-dimensional Parker's equation \citep{1983RPPh...46..973D}
\begin{equation}
\frac{\partial f}{\partial t}-\frac{\partial u}{\partial x}\frac{p}{3}\frac{\partial f}{\partial p}+u\frac{\partial f}{\partial x}=\frac{\partial}{\partial x}\left( \kappa \frac{\partial f}{\partial x} \right) +Q,\label{e1}
\end{equation}
where $ f\left( p,x,t\right)  $ is the isotropic particle distribution function in phase space, $ p $ is the (magnitude of) particle momentum, $ t $ and $ x $ are the temporal and spatial coordinates, $ u\left( x,t\right) $ is the velocity of the background fluid, $\kappa \left( p,x,t\right) $ is the spatial diffusion coefficient of particles, and $ Q\left( p,x,t\right) $ is the source term.
We work in the shock frame, assuming homogeneous background in upstream and downstream of the shock and constant injection at the shock front $ x=0 $, then
\begin{eqnarray}
u\left( x,t\right) &&=u_1+\left( u_2-u_1\right) H\left( x\right) ,\label{e2}\\
\kappa\left( p,x,t\right) &&=\kappa _1\left( p \right) +\left[ \kappa _2\left( p \right) -\kappa _1\left( p \right) \right] H\left( x \right) ,\label{e3}\\
p^2Q\left( p,x,t\right) &&=Q_0\delta \left( p-p_0 \right) \delta \left( x \right) H\left( t \right) ,\label{e4}
\end{eqnarray}
where $ H\left( x \right) $ is the Heaviside step function, $ \delta \left( x \right) $ is the Dirac delta function, and $ p_0 $ is the injection momentum. The subscripts 1 and 2 represent the upstream and downstream, respectively. To simplify the model, we have ignored the shock evolution so that $ u_{1,2} $, $ \kappa_{1,2}\left( p\right) $ and $ Q_0 $ do not vary with time.

For relativistic particles, the diffusion coefficient only depends on the particle rigidity $ R=cp/q $, where $ q $ is the charge of the particle and $ c $ is the speed of light \citep{PhysRevLett.108.081104}.  
For non-relativistic particles, dependence of the diffusion coefficient on $ R $ is complicated due to resonant interactions of particles with kinetic plasma waves \citep{2004ApJ...610..550P}. However, the speed of particles injected into the shock acceleration process should be greater than the shock speed. Particles with lower gyro-frequency, i.e. lower charge-to-mass ratio, should have larger gyro-radii and stronger interaction with plasma waves giving rise to a lower diffusion coefficient \citep{2006ApJ...636..462L} and more efficient acceleration. Observations of CRs do show charge-to-mass ratio dependent characteristics \citep{2008APh....30..133A, 2011ApJ...728..122Y}. We will consider the simple case with $ \kappa=\kappa\left( R_0\right) \left( R/R_0\right) ^\alpha $ and use $ R_0=cp_0/q $ to characterize the charge-to-mass ratio dependence of particle acceleration at relatively low energies. 
Then one may replace $ p $ with $ R $ in the above equations, and Eq. (\ref{e1}) can be readily solved to give $ f\left( R,x,t\right) =f\left( p,x,t\right) {\textrm{d}p}/{\textrm{d}R} $ once one specifies $ u_{1,2},\ \kappa_{1,2}\left( R_0\right) ,\ \alpha_{1,2},\ Q_0 $, and $R_0$ \citep{1991MNRAS.251..340D}. 

To compare with CR observations near the Earth, one also needs to take into account the effects of CR propagation in the Galaxy and heliosphere. For the Milky Way propagation, we use the ``leaky box'' approximation and calculate the propagated flux as
\begin{equation}
J_0\left( R\right) =\frac{H_\textrm{\scriptsize G}^2L_\textrm{\scriptsize S}^2r_\textrm{\scriptsize S}}{V_\textrm{\scriptsize G}D\left( R\right)}
\int_{-\infty}^\infty {\rm d} x f\left( R,x,{t=T_{\textrm{\scriptsize S}}}\right) vp^2,
\end{equation}
where $v$ is the particle speed, $T_\textrm{\scriptsize S}$ is the shock age, $r_\textrm{\scriptsize S}\approx0.03$ yr$^{-1}$ is the mean explosion rate of supernovae in the Galaxy, $L_\textrm{\scriptsize S}$ is the characteristic size of SNR shocks, $H_\textrm{\scriptsize G}$ and $V_\textrm{\scriptsize G}$ are the thickness and volume of the Galaxy, respectively. We adopt $D\left( R\right) =D_0\left[ R/\left( 10\ {\rm GV}\right) \right] ^{1/3}v/c $ according to the boron to carbon flux ratio spectrum of AMS \citep{2016PhRvL.117w1102A}. Such a diffusion coefficient is also consistent with the Kolmogorov spectrum of magnetic field fluctuations in the local interstellar medium \citep{2015ApJ...804L..31B}. Note here we use the spatially integrated spectrum of energetic particles as the CR spectrum injected into the Galaxy by SNRs and $\int_{-\infty}^\infty {\rm d} x \int^\infty_{p_0} {\rm d} p f\left( p,x,{t=T_{\textrm{\scriptsize S}}}\right) p^2=Q_0T_S$.

The force field approximation is adopted to model the solar modulation, and the observed CR flux is given by
\begin{equation}
J\left( R \right) =\frac{vR^2}{v'R'^2}J_0\left( R' \right),
\end{equation}
where 
\begin{equation}
R'^2=R^2+2R\phi \frac{c}{v}+\phi ^2,
\end{equation}
with $ \phi=0.8\ \textrm{GV}$ being an effective potential.

\begin{figure*}[ht]
\centering
\includegraphics[width=1\textwidth]{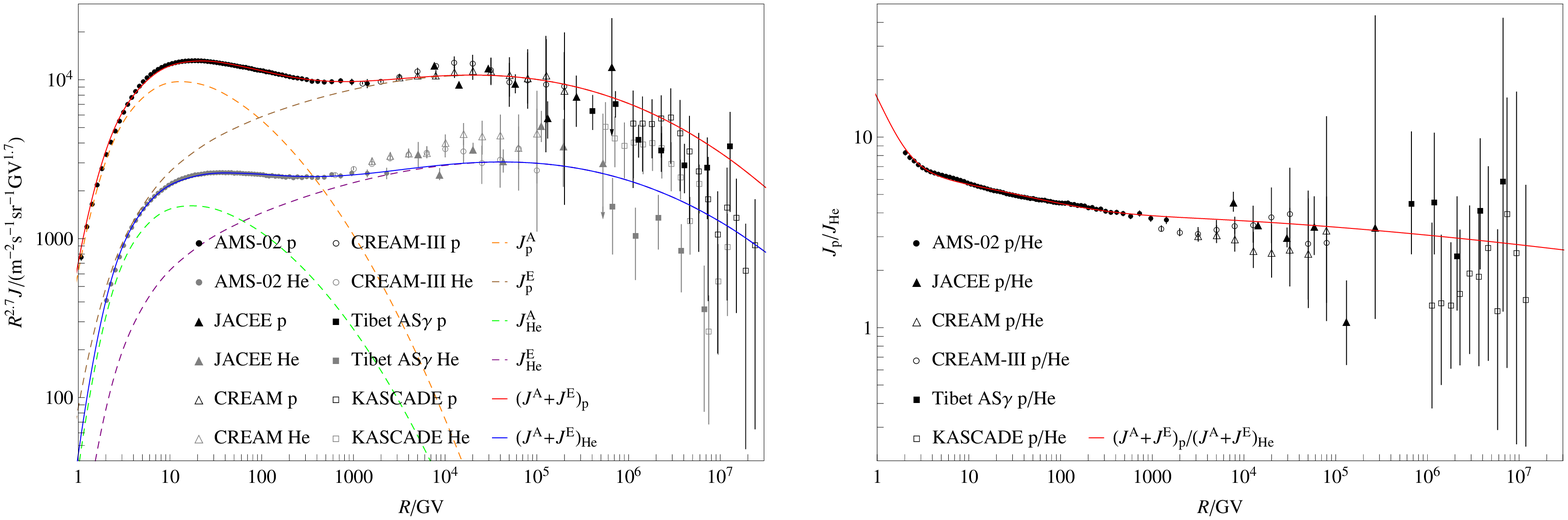}
\caption{
Best fit to the proton and helium spectra (left) and their ratio (right) with a diffusion model described by Eq. (\ref{31}). The data are from AMS \citep{2015PhRvL.114q1103A, 2015PhRvL.115u1101A}, JACEE \citep{1998ApJ...502..278A}, CREAM \citep{2011ApJ...728..122Y}, CREAM-III \citep{2017ApJ...839....5Y}, Tibet AS{$\gamma$} \citep[for HD+SIBYLL;][]{2006PhLB..632...58T}, and KASCADE \citep[for SIBYLL 2.1;][]{2005APh....24....1A}. Except for the AMS data, the p/He flux ratios as functions of rigidity are obtained with spline interpolation of the corresponding energy spectra. The corresponding model parameters are given with the first row of Table \ref{t1}.
\label{f2}}
\end{figure*}
\begin{deluxetable*}{cccccccc}
\tablecaption{Fitting parameters\label{t1}}
\tablecolumns{8}
\tablenum{1}
\tablewidth{0pt}
\tablehead{
\colhead{$ \frac{\kappa_1}{\kappa_2} $} & \vline & \colhead{$ \tau ^{\rm E}_{\rm S} $}&
\colhead{$ \tau ^{\rm A}_{\rm S} $} & \colhead{$ \frac{\left( Q_0 \right)^{\rm E} _{\rm p}}{\left( Q_0 \right)^{\rm E} _{\rm He}} $} & \colhead{$ \frac{\left( Q_0\right)^{\rm A} _{\rm p}}{\left( Q_0 \right)^{\rm A} _{\rm He}} $} & \colhead{$ \frac{\left( Q_0\right)^{\rm E}_{\rm p}}{\left( Q_0\right)^{\rm A}_{\rm p}}\left( \frac{10u_{1}^{\rm A}}{u_{1}^{\rm E}} \right) ^2\frac{\left( \kappa _2L_{\rm S}^{2} \right) ^{\rm E}}{\left( \kappa _2L_{\rm S}^{2} \right) ^{\rm A}}$} & \colhead{$ \frac{\left( Q_0 \right) ^{\rm A}_{\rm p}}{{\rm cm}^{-2}{\rm s}^{-1}{\rm sr}^{-1}}\frac{\kappa _{2}^{\rm A}}{D_0}\left( \frac{L_{\rm S}^{\rm A}}{50\ {\rm pc}} \right) ^2\left( \frac{u_{1}^{\rm A}}{5\times 10^7\ {\rm cm/s}} \right) ^{-2}\left( \frac{V_{\rm G}}{{\rm kpc}^3} \right) ^{-1}\left( \frac{H_{\rm G}}{0.1\ {\rm kpc}} \right) ^2 $}\smallskip
}
\startdata
1 & \vline & 9.0 & 4.7 & 9.1 & 18.5 & 0.2 & $ 8.4\times 10^{-3} $\\
16 & \vline & 10.7 & 6.3 & 9.0 & 17.7 & 0.3 & $ 9.4\times 10^{-4} $
\enddata
\end{deluxetable*}

Observations of SNRs show that the evolution of non-thermal emission associated with the forward shock may be divided into two distinct stages: an early stage (denoted by E) featured with high shock speeds and synchrotron X-ray emission may be associated with the free-expansion and Sedov-Taylor phases of SNRs, and an advanced stage (denoted by A) featured with low shock speed, strong GeV $\gamma$-ray and thermal X-ray emissions implying interaction with molecular clouds may be associated with the late Sedov-Taylor and radiative phases of SNRs \citep{Helder2012, 2017ApJ...834..153Z}. In the following, we use two independent steady strong shocks ($ u_1/u_2=4 $) described by equations (\ref{e1}) $ \sim $ (\ref{e4}) with distinct characteristic speeds and sizes to approximate the time evolution of SNR shocks. A more elaborated model considering details of the shock evolution involves more parameters and may be necessary for study of individual SNRs \citep{2010ApJ...708..965Z}. As will be shown below, this two-phase treatment of SNR shocks is sufficient to explain CR observations. Note that, however, the solution for particle acceleration in each phase is time-dependent. For stage E, we assume a shock speed of $u_1^{\rm E}\sim 10^9$ cm/s. All particles are further assumed to be injected at $v_0=10^9$ cm/s. The corresponding critical rigidity is
\begin{equation}
\left( R_0 \right) _{\textrm{\scriptsize He}}^{\textrm{\scriptsize E}}=2\left( R_0 \right) _{\textrm{\scriptsize p}}^{\textrm{\scriptsize E}}\approx \frac{0.938}{15}\ \textrm{GV}.\label{28}
\end{equation}
For stage A, the shock speed and particle injection velocity are smaller by an order of magnitude.

For each of these two stages, we define a dimensionless shock age as $\tau_\textrm{\scriptsize S}= T_\textrm{\scriptsize S}/T\,[\left( R_0\right)_{\rm p}] $, where $ T\left( R\right) $ is the acceleration timescale at $ R $ \citep{1983RPPh...46..973D}
\begin{equation}
T\left( R \right) =\frac{4}{u_1-u_2}\left[ \frac{\kappa _1\left( R\right)}{u_1}+\frac{\kappa _2\left( R\right)}{u_2} \right]. \label{time_R0}
\end{equation}
One can adjust $\tau_\textrm{\scriptsize S}$, $\kappa$, and $Q_0$ to fit the observed CR spectra.
\section{Results}\label{sec3}
We first consider a relatively simpler case with
\begin{equation}
\frac{\kappa _1}{\kappa _2}=1,\ \ \alpha _1=\alpha _2=0,\label{31}
\end{equation}
which may correspond to diffusion dominated by turbulent convection \citep{1993PhyU...36.1020B}. There are therefore six main parameters to fit the observed proton and helium spectra. The best fit to the AMS spectra and CREAM proton spectrum are shown in Figure \ref{f2}, and the fitting parameters are shown in Table \ref{t1}. We see that for shock ages many times higher than the particle acceleration timescale, the time-dependent effect of particle acceleration process can still be important. Since protons and helium are injected for acceleration with different rigidities (Eq. (\ref{28})), the ratio of injection rates of protons and helium is not identical to the background abundance. A lower proton to helium ratio in the early stage required to fit the data implies that the metallicity of background in such a stage is higher than that in the advanced stage.

\begin{figure*}[ht]
\centering
\includegraphics[width=1\textwidth]{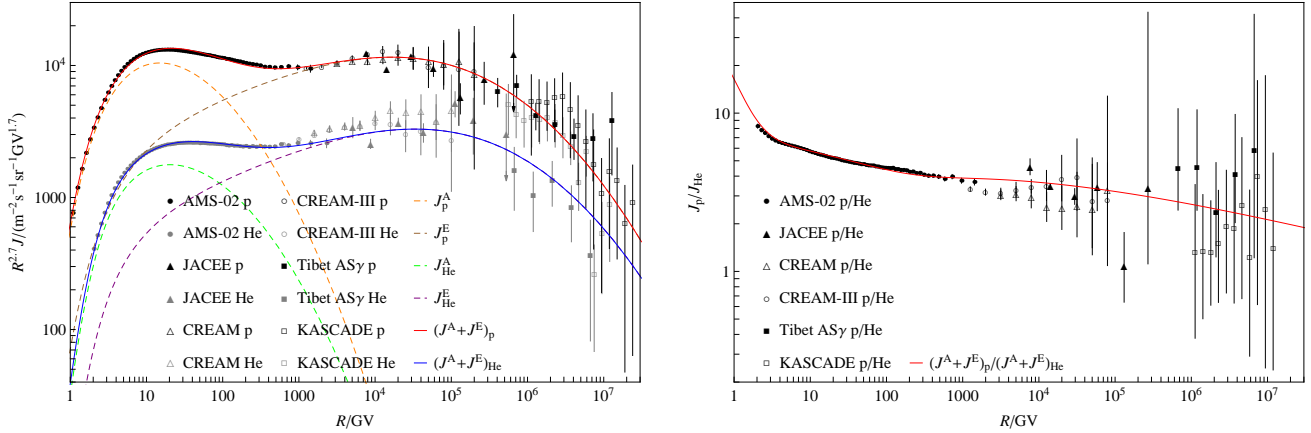}
\caption{
Same as Figure \ref{f2} but using Eq. (\ref{35}), and corresponding model parameters are given with the second row of Table \ref{t1}.\label{f3}
}
\end{figure*}

We notice that the model slightly over-produces CR fluxes at high rigidities (above the ``knee'' of $\sim 10^5$ GV). Considering the fact that the turbulence (or fluctuating magnetic field) in downstream of the shock should be stronger than that in upstream, we expect $ \kappa_1/\kappa_2\gg 1 $. Adopting \citep[for this case there is an analytic solution, see][]{1991MNRAS.251..340D}
\begin{equation}
\frac{\kappa _1}{\kappa _2}=16,\ \ \alpha _1=\alpha _2=0,\label{35}
\end{equation}
as shown in Figure \ref{f3}, the CR spectral fit is improved, and the fitting parameters are also shown in Table \ref{t1}.

From Eq. (\ref{time_R0}) one can derive the diffusion coefficient in downstream of the shock as
\begin{equation}
\kappa _2\approx \frac{5.9\times 10^{25}}{4+\kappa _1/\kappa _2}\frac{T_{\textrm{\scriptsize S}}}{\tau _{\textrm{\scriptsize S}}\ \textrm{kyr}}\left( \frac{u_1}{10^8\ \textrm{cm/s}} \right) ^2\ \textrm{cm}^2\textrm{s}^{-1}.\label{e5}
\end{equation}
For characteristic ages and sizes of SNRs in the two stages mentioned above
\begin{eqnarray}
&&T_\textrm{\scriptsize S}^\textrm{\scriptsize E}\sim\textrm{kyr},\ \ 
L_\textrm{\scriptsize S}^\textrm{\scriptsize E}=\left( T_\textrm{\scriptsize S}u_1\right) ^\textrm{\scriptsize E}\sim 5
\left(\frac{u_1^{\textrm{\scriptsize E}}}{5\times 10^8\ \textrm{cm/s}} \right) \ \textrm{pc},\nonumber\\
&&T_\textrm{\scriptsize S}^\textrm{\scriptsize A}\sim 100\ \textrm{kyr},\ \
L_\textrm{\scriptsize S}^\textrm{\scriptsize A}=\left( T_\textrm{\scriptsize S}u_1\right) ^\textrm{\scriptsize A}\sim 50
\left(\frac{u_1^{\textrm{\scriptsize A}}}{5\times10^7\ \textrm{cm/s}} \right) \ \textrm{pc},\ \ \ \ \ \ \ \ \label{e6}
\end{eqnarray}
and typical CR propagation parameters in the Galaxy
\begin{equation}
V_\textrm{\scriptsize G}\sim \textrm{kpc}^3,\ \ H_\textrm{\scriptsize G}\sim 0.1\ \textrm{kpc},\ \ D_0\sim 10^{29}\ \textrm{cm}^2\textrm{s}^{-1},\label{e7}
\end{equation}
inserting $ \tau_{\rm S} $ given in Table \ref{t1} into Eq. (\ref{e5}), we can derive $ \kappa $ and $ Q_0 $, and the results are shown in Table 2. 
We see that the diffusion coefficient $\kappa \sim 0.01 uL_{\rm S}$ and $uL_{\rm S}\sim D\left( {\rm GV}\right) $, which are quite reasonable. Compared to the case $ \kappa_1/\kappa_2=1 $, slightly lower values of diffusion coefficients are inferred for $ \kappa_1/\kappa_2=16 $ to compensate the lower level of turbulence assumed in the upstream.

\begin{deluxetable*}{cccccc}
\tablecaption{Derived diffusion coefficients and injection rates\label{t2}}
\tablecolumns{6}
\tablenum{2}
\tablewidth{0pt}
\tablehead{
\colhead{$ \frac{\kappa_1}{\kappa_2} $} & \vline & \colhead{$ \frac{\kappa _2^{\rm E}}{{\rm cm}^2{\rm s}^{-1}}\left( \frac{u_1^{\rm E}}{5\times 10^8\ {\rm cm/s}} \right) ^{-1}\left( \frac{L_{\rm S}^{\rm E}}{5\ {\rm pc}} \right) ^{-1} $}&
\colhead{$ \frac{\kappa _2^{\rm A}}{{\rm cm}^2{\rm s}^{-1}}\left( \frac{u_1^{\rm A}}{5\times 10^7\ {\rm cm/s}} \right) ^{-1}\left( \frac{L_{\rm S}^{\rm A}}{50\ {\rm pc}} \right) ^{-1} $} & \colhead{$ \frac{\left( Q_0\right) _{\rm p}^{\rm E}}{{\rm cm}^{-2}{\rm s}^{-1}{\rm sr}^{-1}}\frac{T_{\rm S}^{\rm E}}{{\rm kyr}}\left( \frac{L_{\rm S}^{\rm E}}{5\ {\rm pc}} \right)^2 $} & \colhead{$ \frac{\left( Q_0\right) _{\rm p}^{\rm A}}{{\rm cm}^{-2}{\rm s}^{-1}{\rm sr}^{-1}}\frac{T_{\rm S}^{\rm A}}{{\rm 100\ kyr}}\left( \frac{L_{\rm S}^{\rm A}}{50\ {\rm pc}} \right)^2$}\smallskip
}
\startdata
1 & \vline & $ 3.3\times 10^{25} $ & $ 6.4\times 10^{25} $ & 570 & 10\\
16 & \vline & $ 6.9\times 10^{24} $ & $ 1.2\times 10^{25} $ & 440 & 8
\enddata
\end{deluxetable*}

The density of particles injected for acceleration can be estimated by
$ Q_0\sim nv_0/\left( 4\pi \right)$, which turns out to be
\begin{equation}
n^{\textrm{\scriptsize E}} _{\textrm{\scriptsize p}}\sim 10^{-5} \ \textrm{cm}^{-3},\ \
n^{\textrm{\scriptsize A}} _{\textrm{\scriptsize p}}\sim 10^{-6} \ \textrm{cm}^{-3}.
\end{equation}
Both values are several orders of magnitude lower than densities of the background plasmas. Since the relativistic particle distribution is very soft with an index greater than 2.3 and non-relativistic particle momentum distribution approaches the state-steady with an index of 2, the total energy of CRs injected into the medium by an SNR is estimated as $E\sim 4\pi \left( Q_0qR_0/2\right) _{\rm p}L_{\rm S}^2 T_{\rm S}$, which is on the order of $10^{48}$ erg and $10^{50}$ erg for the early and advanced stages, respectively, justifying the linear treatment of diffusive shock acceleration. The bulk of CRs is therefore accelerated in dense medium by relatively slower shocks.
\section{Conclusion and Discussion}\label{sec4}
Since the discovery of anomalous fine structures in the energy spectra of CRs, there have been extensive investigations focusing on CR acceleration and propagation processes. Here we show that, considering time evolution of the linear diffusive shock acceleration process, the observed rigidity dependence of proton to helium flux ratio may just suggest that particle diffusion process near shock front of CR accelerators is dominated by turbulence convection giving rise to a diffusion coefficient weakly dependent on the particle rigidity. Recent TeV observations of SNR RX J1713.7-3946 do support such a scenario \citep{2016arXiv160908671H}. In this paper, we only consider cases with $ \alpha=0 $. For $ \alpha_1=\alpha_2=1/30 $ with $ \kappa_1/\kappa_2=1 $, we can get spectra similar to the second model. For even higher values of $\alpha$, the time-dependent particle distribution approaches to the steady-state spectrum at low energies and cuts off too sharply at the energy where the acceleration timescale is comparable to the shock age to explain the CR spectra near the ``knee''. The proton to helium flux ratio will be constant at low energies, similar to the two component model proposed by \citet{2015ApJ...815L...1T} and the rigidity dependence of $ D $ needs to be adjusted to fit the observed CR spectra below the ``knee''.

The observed CR spectral hardenings near $\sim 200$ GV may be attributed to two stages of the SNR evolution. In the early free expansion and Sedov-Taylor stage, the shock speed and background metallicity are high, and the acceleration dominates the CR fluxes above $\sim$ 200 GV. In the advanced radiative stage, the shock is propagating in dense medium slowly, giving rise to a softer spectrum and higher proton to helium ratio. These two stages of SNRs are actually commonly seen in multi-wavelength observations \citep{Helder2012, 2017ApJ...834..153Z}. Our model therefore links the observed CR spectral anomalies to multi-wavelength observations of SNRs, implying the dominance of Galactic CR acceleration by SNRs. In the paper, we adopt characteristic parameters for isolated SNRs. The model can also be applied to CR acceleration in super bubbles \citep{Ohira16}.
 
The model has a soft spectrum (with an index of $\sim 2.4$) of energetic particles  injected into the Galaxy by SNRs, which will produce a lower level of CR anisotropy than steady-state diffusive shock models (usually with an injection index of $\sim 2$). It also predicts a gradual softening of the spectra at high energies, which may be responsible for the ``knee'' of CR spectra. Future observations of the spectra by e.g., LHAASO may be useful in testing this model prediction. 

Time-dependent stochastic particle acceleration by turbulent plasma waves in the shock downstream can produce similar results \citep{2006ApJ...647..539B,2010MNRAS.406.1337F}. In this case energy dependence of the acceleration timescale needs to be weak and the flux of particles escaping from SNRs during acceleration, which is an essential element of stochastic particle acceleration mechanism, can also be obtained. The time-dependent solutions may also explain the hardening of SNR radio spectrum with age \citep{Reynolds2012, 2017ApJ...834..153Z}. More detailed modelling and comparison with SNR observations may be able to distinguish these different particle acceleration scenarios.
\acknowledgments
This work is supported in part by the National Natural Science Foundation
of China (Nos. 11173064, 11233001, and 11233008), and the 100 Talents 
program of Chinese Academy of Sciences.

\end{document}